\documentclass[epj]{svjour}

\usepackage{graphicx}
\usepackage{fancyhdr}
\def\makeheadbox{{%  remove EPJC header
 }}
\def\mytitle{My title} 
\def\myauthors{My name}  
\def\mytype{My type of session}
\def\mysession{My session}
\def\mytitle{Orbifolds versus smooth compactifications} %Put your title here!
\def\myauthors{Gabriele Honecker}    %Put your name here!
\def\mytype{Contributed Talk}    
\def\mysession{Theoretical Models}

\begin{document}
\title{Orbifolds versus smooth heterotic compactifications}
\author{Gabriele Honecker
\thanks{\emph{Email:} Gabriele.Honecker@cern.ch}%
}                     % Do not remove
\institute{PH-TH Division, CERN, 1211 Geneva 23, Switzerland
}
%
%\date{Received: date / Revised version: date}}
\date{September 13, 2007}
\abstract{
Following the recent exploration of smooth heterotic compactifications with 
unitary bundles, orbifold compactifications in six dimensions can be 
shown to correspond in the blow-up to compactifications with U(1) gauge backgrounds.
A powerful tool is the comparison of anomaly polynomials.
The presentation here focuses on heterotic $SO(32)$ compactifications in six dimensions 
including five-branes. Four dimensional and $E_8 \times E_8$ models are briefly commented on.
 % end of PACS codes
} %end of abstract
\maketitle
\section{Introduction}
\label{intro}

Orbifold compactifications of the heterotic string on the one hand 
have been employed for the past twenty years, and on the other hand heterotic 
compactifications with $SU(n)$ ($n=4,5$)
 bundles on smooth manifolds have developed in the last ten years.
These two seemingly unrelated approaches are shown to be closely related
when the orbifold blow-ups are interpreted as smooth compactifications with
$U(1)$ bundles instead of the $SU(n)$ bundles which have been the 
main focus for model building. For $T^4/Z_N$ embeddings with $N=2,3$,
there is a direct identification of the orbifold shift vector with
the embedding of a line bundle $L$~\cite{Honecker:2006qz},
\begin{equation}
\label{eq:Shift-Line-Ident}
\frac{1}{N}(1_{n_1},2_{n_2},\ldots,0_{n_0}) 
\rightarrow (L_{n_1}, L^2_{n_2},\ldots,0_{n_0}),
\end{equation}
where the lower index $n_i$ denotes the number of identical entries $i$ in 
the shift vector and $\sum_i n_i = 16$.
Furthermore, compactifications with $U(n)$ bundles amplify the possibilities
to obtain four dimensional vacua with standard model or GUT gauge groups
and are S-dual to Type I compactifications with non-Abelian bundles 
on D9-branes.

The present exposition is for concreteness  focused on $SO(32)$ 
heterotic compactifications in six dimensions. Four dimensional cases and 
$E_8 \times E_8$ compactifications are briefly commented on.

\section{Heterotic $T^4/Z_N$ Orbifolds}
\label{sec:Orbifolds} 

Abelian $T^{2n}/Z_N$ orbifolds of the heterotic string are 
described by two shift vectors, the space-time shift $\vec{v}$ 
which encodes a $Z_N$ rotation $z_j \rightarrow e^{2 \pi i v_j} z_j$
on $j=1 \ldots n$ complex coordinates, and a gauge shift $\vec{V}$
which embeds the orbifold action in the gauge degrees of freedom.
The vectorial shift vectors for $n$ complex compact dimensions and 
embeddings in $SO(32)$ are given by
\begin{equation}
\vec{v} = \frac{1}{N}(\sigma_1 \ldots \sigma_n),
\quad
\vec{V} = \frac{1}{N}(\Sigma_1 \ldots \Sigma_{ 16}),
\end{equation}
with $\sigma_j, \Sigma_k$ integer and $\sum_j \sigma_j =0$ to
ensure that the space-time orbifold is the singular limit of 
a Calabi-Yau $n$-fold. In order to obtain 
supersymmetric models, two stringy constraints 
have to be met. These are on the one hand the quadratic `level-matching' condition,
which ensures the modular invariance of the partition function
and mixes space-time and gauge shifts, and on the other hand a 
linear condition on the gauge shift ensuring the existence of spinors in the gauge 
bundle,
\begin{equation}
\label{eq:Orbifold_Constraints}
N \sum_i \bigl(V_i^2 - v_i^2 \bigr) = 0 \, {\rm mod} \, 2,
\quad
N \sum_i V_i = 0 \, {\rm mod} \, 2.
\end{equation}
The massless spectrum for a given choice of shift vectors in $SO(32)$ embeddings
is obtained as follows: those $SO(32)$ weight vectors $\vec{w}=(\underline{\pm 1, \pm 1, 0_{14}})$ with $\vec{w} \cdot \vec{V} \in Z$ provide the non-Abelian generators 
of the gauge group and those with $\vec{w} \cdot \vec{V} \notin Z$ the 
untwisted matter states. The total rank of the gauge group is 16, and depending 
on the chosen gauge shift the gauge group can contain several $U(1)$ factors.

The twisted spectrum consists of $n=1 \ldots N-1$ sectors with twisted 
ground states obtained from $\vec{w} - n \vec{V}$. Oscillators lift the
tachyionic vacuum to the massless level, and multiplicities are obtained from
the counting of space-time fixed points. For the $T^4/Z_N$ orbifolds with $\vec{v}=\frac{1}{N}(1,-1)$ these are 16 $Z_2$ fixed points for $N=2$, nine $Z_3$ fixed 
points for $N=3$ or four $Z_4$ and 16 $Z_2$ fixed points for $N=4$ 
and one $Z_6$, nine $Z_3$ and 16 $Z_2$ fixed points for $N=6$. As an
example, the spectra for the `standard embeddings' with $\vec{V}=\frac{1}{N}(1,1,0^{14})$ are listed in table~\ref{tab:Orbifold_SE}.
The gauge group is $SO(28) \times SU(2)^2$ for $N=2$ and $SO(28) \times SU(2) \times
U(1)$ otherwise. A counting of non-Abelian degrees of freedom 
yields $10 ({\bf 28},{\bf 2}) + 66 ({\bf 1})$ for all $N$ 
with identical $U(1)$ charge assignments in the untwisted sectors, but 
different $U(1)$ charges $1 + \frac{m}{n}$ in the $n^{th}$ twisted sector
and $m$ integer.
\begin{table}[htb]
\caption{Spectra of $SO(32)$ $T^4/Z_N$ standard embeddings.}
\label{tab:Orbifold_SE}    
\begin{tabular}{lll}
\hline\noalign{\smallskip}
 & N=2 & N=3  \\
\noalign{\smallskip}\hline\noalign{\smallskip}
$\theta^0$ & $({\bf 28},{\bf 2},{\bf 2}) + 4 ({\bf 1})$
& $({\bf 28},{\bf 2})_1 + 2 ({\bf 1})_0 + ({\bf 1})_2$ 
\\
$\theta^1$ & $8({\bf 28},{\bf 1},{\bf 2})+ 32({\bf 1},{\bf 2},{\bf 1})$
& $\left\{ \begin{array}{c}  9({\bf 28},{\bf 2})_{\frac{1}{3}} + 45({\bf 1})_{\frac{2}{3}}
\\ + 18 ({\bf 1})_{\frac{4}{3}}\end{array}\right.$
\\
\hline\noalign{\smallskip}
 & N=4 & N=6  \\
\noalign{\smallskip}\hline\noalign{\smallskip}
$\theta^0$ & $({\bf 28},{\bf 2})_1 + 2 ({\bf 1})_0 $ 
& $({\bf 28},{\bf 2})_1 + 2 ({\bf 1})_0$ 
\\
$\theta^1$ & $\left\{ \begin{array}{c} 4({\bf 28},{\bf 2})_{\frac{1}{2}} + 24 ({\bf 1})_{\frac{1}{2}}\\ + 8  ({\bf 1})_{\frac{3}{2}}
\end{array}\right. $
&$({\bf 28},{\bf 2})_{\frac{2}{3}} + 8 ({\bf 1})_{\frac{1}{3}} + 2 ({\bf 1})_{\frac{5}{3}}$ 
\\
$\theta^2$ & $ 5 ({\bf 28},{\bf 2})_0 + 32 ({\bf 1})_1$ &
$\left\{ \begin{array}{c}  5({\bf 28},{\bf 2})_{\frac{1}{3}} + 22 ({\bf 1})_{\frac{2}{3}}\\ + 10 ({\bf 1})_{\frac{4}{3}}\end{array}\right. $
\\
$\theta^3$ & - & $3 ({\bf 28},{\bf 2})_0 + 22 ({\bf 1})_1$  
\\
\noalign{\smallskip}\hline
\end{tabular}
\end{table}

For any given six dimensional spectrum, field theory anomalies arising from fermions and tensors running in loops
can be computed.
Using the complete list of $T^4/Z_N$ spectra for $N=2,3,4$ and some $N=6$ 
examples, the anomaly polynomial for $SO(32)$ heterotic orbifold compactifications takes the 
form~\cite{Honecker:2006qz}
\begin{eqnarray}
&& I_8 =
\bigl( {\rm tr} R^2 + \sum_i \alpha_i {\rm tr}_{SO(2M_i)} F^2 +
\sum_j \beta_j {\rm tr}_{SU(N_j)} F^2
\bigr.
\nonumber\\
&&\bigl.
+\sum_k \gamma_k F_{U(1)_k}^2 
+ \sum_{i<j} \delta_{ij} \,  F_{U(1)_i} \, F_{U(1)_j}
\bigr) \times
\bigl({\rm tr} R^2 -
\bigr.
\nonumber\\
&&\bigl.
- \sum_i  {\rm tr}_{SO(2M_i)} F^2  -2 \, \sum_j  {\rm tr}_{SU(N_j)} F^2
+\sum_k \tilde{\gamma}_k F_{U(1)_k}^2
\bigr)
\label{Eq_Anomaly_Orbifold_SO32}
\end{eqnarray}
with $\alpha_0 =2$ for all $SO(2M_0)$ gauge groups with fundamental
representations only, $\alpha_1 \leq 1$ otherwise, $\beta_j \leq 2$
and nearly always even, $\gamma_k< -7$ and 
$\tilde{\gamma}_k  \leq -2$ with the latter always even.
The $T^4/Z_3$ orbifold admits at most one $SO(2M_0)$ gauge factor 
associated with $M_0$ zero entries in the gauge shift $\vec{V}$ 
and $\alpha_0 =2$, whereas the $T^4/Z_N$ orbifolds with $N$ even 
admit a second $SO(2M_1)$ gauge factor associated with the entries 
$\Sigma_{i_1}= \ldots =\Sigma_{i_{M_1}} =N/2$ with spinorial
representations in the twisted spectrum and $\alpha_1 \leq 1$.   
The anomaly polynomial at the orbifold point factorises completely
into $4 \times 4$, and as discussed below for the smooth $K3$ 
compactifications, this signals the fact that $U(1)$ factors at the 
orbifold point are massless.

The scalar potential in six dimensions is completely determined by
D-term interactions,
\begin{equation}\label{eq:6D_potential}
V=\sum_{a,\alpha} D^{a,\alpha} D^{a,\alpha}
\quad
{\rm with}
\quad
 D^{a,\alpha} = \Phi^{\dagger}_i \sigma^a t^{\alpha}_{ij} \Phi_j,
\end{equation}
with the Pauli matrices $\sigma^a$, generators of the gauge groups $t^{\alpha}_{ij}$
and matter fields $\Phi_j$.

The anomaly polynomials and gauge shift vectors admit by comparison
an interpretation of the orbifold models as smooth compactifications 
with $U(1)$ bundles as discussed in the next section provided that  twisted scalars receive 
vacuum expectation values along flat directions of the scalar 
potential~(\ref{eq:6D_potential}) thereby blowing up the singularities and 
breaking the orbifold point $U(1)$.

\section{The heterotic string on K3}
\label{sec:HeteroticK3} 

Smooth compactifications of the heterotic string on Calabi-Yau $n$-folds 
require for the gauge bundle $\bar{F} = \oplus_i \bar{F}_i$  to preserve
supersymmetry that each component $\bar{F}_i$ is a holomorphic (1,1)-form satisfying 
at tree level the primitivity condition 
$ \int_{CY_n} J^{n-1} \wedge {\rm tr} \bar{F}_i = 0$.
On $K3=CY_2$ the latter condition is exact, whereas on $CY_3$ the primitivity
condition receives a 1-loop correction.
Furthermore, consistent compactifications require the Bianchi identity
on the 3-form $H=dB -\frac{\alpha'}{4}(\omega_{YM}-\omega_{L})$ which
is quadratic in the gauge bundle to be satisfied as well as the K-theory 
constraint which is linear in the bundle,
\begin{equation}
\label{eq:CC_smooth}
{\rm tr} \bar{F}^2 - {\rm tr} \bar{R}^2 =0,
\quad
(2 \pi)^{-1} {\rm tr} \bar{F} \in H^2(CY_n,2 Z).
\end{equation}
Eq.~(\ref{eq:CC_smooth}) can be directly compared to its orbifold 
counterpart~(\ref{eq:Orbifold_Constraints}).
The massless spectrum is obtained by decomposing the adjoint representation of 
$SO(32)$,
\begin{equation}
{\bf 496} \to
\left(\begin{array}{c}
({\bf Anti}_{SO(2M)}) \\
\sum_j ({\bf Adj}_{U(N_j)};{\bf Adj}_{U(n_j)})\\
\sum_j ({\bf Anti}_{U(N_j)};{\bf Sym}_{U(n_j)}) + c.c\\
\sum_j              ({\bf Sym}_{U(N_j)};{\bf Anti}_{U(n_j)}) + c.c.\\
\sum_{i < j} ({\bf N}_i,{\bf N}_j;{\bf n}_i,{\bf n}_j) + ({\bf N}_i,\bar{{\bf N}}_j,{\bf n}_i,\bar{{\bf n}}_j) + c.c. \\
\sum_j ({\bf 2M}, {\bf N}_j;{\bf n}_j) + c.c.\\
\end{array}\right),
\end{equation}
and embedding $U(n_j)$ bundles $V_j$ inside $U(N_j n_j)$ factors. The gauge group 
is $SO(2M) \times \prod_j U(N_j)$ with the massless representations counted by 
cohomology classes of the associated bundles as listed in table~\ref{TchiralSO32}.
\begin{table}[htb]
\caption{Massless spectra in terms of cohomology classes.}
\label{TchiralSO32}
\begin{tabular}{ll}
\hline\noalign{\smallskip}
reps. & $SO(2M) \times \prod_i U(N_i)$   \\
\noalign{\smallskip}\hline\noalign{\smallskip}
$({\bf Adj}_{U(N_i)})_{0(i)}$ & $H^*(CY_n,V_i \otimes V_i^{\ast})$  \\
$({\bf Sym}_{U(N_i)})_{2(i)}$ & $H^*(CY_n,\bigwedge^2 V_i)$  \\
$({\bf Anti}_{U(N_i)})_{2(i)}$ & $H^*(CY_n, \bigotimes^2_s  V_i)$  \\
$({\bf N}_i,{\bf N}_j)_{1(i),1(j)}$ & $H^*(CY_n, V_i \otimes V_j)$ \\
$({\bf N}_i,\bar {\bf N}_j)_{1(i),-1(j)} $ &  $H^*(CY_n, V_i \otimes V_j^{\ast})$ \\
$({\bf Adj}_{SO(2M)})_0$ & $H^*(CY_n,{\cal O})$ \\
$({\bf 2M}, {\bf N}_i)_{1(i)}$ & $H^*(CY_n, V_i)$\\
\noalign{\smallskip}\hline
\end{tabular}
\end{table}
The chiral part of the spectrum is computed from the Euler
characters of the bundles $W$,
\begin{eqnarray}
\chi(CY_n, W) 
&=& \sum_{j=0}^n (-1)^j {\rm dim} H^j(CY_n,W) 
\nonumber\\
&=& \int_{CY_n} {\rm ch}(W) {\rm Td}(CY_n),
\end{eqnarray}
with  the Chern characters  ${\rm ch}_k (W) = (k! (2\pi)^k)^{-1} {\rm tr} \bar{F}^k$
of the bundles and  the Todd 
class of the manifold ${\rm Td}(CY_n) = 1 + \frac{1}{12} c_2(CY_n) + \ldots$.
For $K3=CY_2$, the index $\chi(K3,W) = {\rm ch}_2 (W) + 2 \, {\rm rank}(W)$
actually counts the number of vector minus the number of hyper multiplets,
and the complete massless spectrum is easily computed for a given embedding.  

The $U(1)$s in smooth compactifications generically become massive through 
the generalised Green-Schwarz mechanism involving antisymmetric tensor modes 
on the $CY_n$, namely in terms of the expansion of the ten dimensional dual 6-form
$B^{(6)} = \ell_s^{2n-2} b^{(8-2n)}_k \sum_k \widehat{\omega}_k +\ldots$ along  
$(2n-2)$ forms $\widehat{\omega}_k$ on $CY_n$, the mass terms arising from wrapped antisymmetric tensor modes
are given by 
\begin{equation}\label{eq:GS_mass}
S_{mass} \sim \ell_s^{2n-8}\sum_k \sum_i \int_{M^{10-2n}} b^{(8-2n)}_k \wedge [{\rm tr} F_i\bar{F}_i]^k,
\end{equation}
i.e. the $U(1)$ masses are of the order of the string scale.
For $K3$ compactifications, this is a complete set of mass terms, whereas on $CY_3$-folds,
also the coupling to the unwrapped antisymmetric tensor mode contributes.

As an example, consider a line bundle $L$ embedded in $U(2)$, also
denoted as $(L,L,0_{14})$,
with ${\rm ch}_2(L)=-12$ in order to fulfill the Bianchi identity.
The gauge group is $SO(28) \times SU(2)_{\times U(1)_{massive}}$
with matter spectrum $10({\bf 28},{\bf 2})_1 + 46({\bf 1},{\bf 1})_2$
and twenty neutral hyper multiplets parameterising the $K3$ geometry.
The counting of non-Abelian representations agrees with that of the
orbifold `standard embeddings' in section~\ref{sec:Orbifolds}.

For the most general embedding of $U(n_i)$ bundles in $SO(32)$, the 
anomaly polynomial in six dimensions is given by~\cite{Honecker:2006dt}
\begin{eqnarray}
&& I_8 = \frac{1}{3} \bigl(\sum_i c_1(V_i) {\rm tr}_{U(N_i)} F\bigr) \times
\label{eq:I_8_smooth}\\
&&
\quad \times  \bigl(\sum_j c_1(V_j) \bigl[
{\rm tr}_{U(N_j)} F \, {\rm tr} R^2 - 16  {\rm tr}_{U(N_j)} F^3
\bigr]\bigr)
\nonumber\\
&& + \bigl({\rm tr} R^2 +2 {\rm tr}_{SO(2M)} F^2 + 4 \sum_i ({\rm ch}_2(V_i) +n_i) {\rm tr}_{U(N_i)} F^2
\bigr)
\nonumber\\
&& \quad \times  \bigl({\rm tr} R^2 - {\rm tr}_{SO(2M)} F^2 -2 \, \sum_i \, n_i \, {\rm tr}_{U(N_i)} F^2
\bigr).
\nonumber
\end{eqnarray}
The anomaly eight-form~(\ref{eq:I_8_smooth}) on $K3$ factorises as $ 2 \times 6 + 4 \times 4$, 
where the `2' arises from those Green-Schwarz couplings~(\ref{eq:GS_mass}) providing
$U(1)$ masses.

Comparing the coefficients of the non-Abelian gauge factors in
the last two lines of ~(\ref{eq:I_8_smooth}) with those at the orbifold 
point~(\ref{Eq_Anomaly_Orbifold_SO32}) 
leads to 
$\alpha_i \stackrel{!}{=} 2$, $\beta_j \stackrel{!}{=} 4({\rm ch}_2(V_j) + n_j)$, $-2 \stackrel{!}{=}-2n_j$. 
The last condition, $n_j=1$, reveals 
that orbifold models correspond in the blown-up phase to smooth embeddings 
with $U(1)$ bundles. The smooth instanton numbers ${\rm ch}_2(V_j)$ 
can then be computed using the second identification, and finally $\alpha_0=2$
is fulfilled for those $SO(2M_0)$ gauge factors at the orbifold point arising
from zeros in the shift vector. In the other cases, the smooth gauge group is 
only $SU(M_1) \subset SO(2M_1)$, and spinorial representations decompose 
into singlets and antisymmetric representations of $SU(M_1)$.  

The primitivity condition $\int_{CY_n} J^{n-1} \wedge {\rm tr} \bar{F}_i =0$ at  tree level 
is at the orbifold point trivially fulfilled since the gauge bundle is localised
at fixed points whose exceptional divisors have zero volume. More 
results on the blowing-up procedure are given in~\cite{Nibbelink:2007rd,Nibbelink:2007pn}.

\section{Including five-branes}
\label{sec:H5-branes}

It is straightforward
to include some non-perturbative objects, the heterotic 5-branes.
For $SO(32)$ compactifications, $N_a$ coincident 5-branes provide a $Sp(2N_a)$
gauge factor, and the matter spectrum in table~\ref{TchiralSO32} is extended by antisymmetric 
and bifundamental states counted by extensions rather than cohomologies as listed in table~\ref{TchiralSO32_H5}.
\begin{table}[htb]
\caption{Massless states from 5-branes counted by extension groups.}
\label{TchiralSO32_H5}
\begin{tabular}{ll}
\hline\noalign{\smallskip}
reps. & $SO(2M) \times \prod_{i} U(N_i) \times \prod_{a} Sp(2N_a)$   \\
\noalign{\smallskip}\hline\noalign{\smallskip}
$({\bf Anti}_{Sp(2N_a)})$ & ${\rm Ext}^*_{CY_n}({\cal O}|_a,{\cal O}|_a)$
\\
$({\bf N}_i,{\bf 2N}_a)_{1(i)}$ & ${\rm Ext}^*_{CY_n}(V_i,{\cal O}|_a)$
\\
$({\bf 2M},{\bf 2N}_a)$ & ${\rm Ext}^*_{CY_n}({\cal O},{\cal O}|_a)$
\\
$({\bf 2N}_a,{\bf 2N}_b)$ & ${\rm Ext}^*_{CY_n}({\cal O}|_a,{\cal O}|_b)$
\\
\noalign{\smallskip}\hline
\end{tabular}
%\vspace*{1cm} 
\end{table}
The sky-scraper sheafs ${\cal O}|_a$ describing the 5-branes have ${\rm ch} ({\cal O}|_a)=
(0,0,-\gamma_a,0)$ where $\gamma_a=1$ for a 5-brane which is point like
on $K3$ and $\gamma_a$ the Poincar\'e dual 4-form of a 5-brane wrapping a 2-cycle
on $CY_3$.
The Bianchi identity is modified to 
\begin{equation}
{\rm tr} \bar{F}^2 - {\rm tr} \bar{R}^2 - 16 \pi^2 N_a \gamma_a=0,
\end{equation}
and the supersymmetry conditions on the bundles are unchanged. The anomaly 
polynomial~(\ref{eq:I_8_smooth}) 
receives in the third line an additional 
term $-2 {\rm tr}_{Sp(2N_a)} F^2$~\cite{Honecker:2006dt}.
On $K3$, bifundamental representations of 5-branes at different points are 
massive with the mass proportional to their distance. There is one hyper multiplet
in the antisymmetric representation of $Sp(2N_a)$, $n_i$ hyper multiplets
transforming as  $({\bf N}_i,{\bf 2N}_a)_{1(i)}$ 
and a half-hyper multiplet in the $({\bf 2M},{\bf 2N}_a)$ representation. 

As an example, take again a line bundle $L$ embedded in $U(2)$, but this time with 
${\rm ch}_2(L) = -3$. The Bianchi identity is fulfilled in the presence of 18 5-branes.
The gauge group is $SO(28) \times SU(2)_{\times U(1)_{massive}} \times Sp(36)$,
and the matter spectrum consists of $({\bf 28},{\bf 2}_1;{\bf 1}) + ({\bf 1},{\bf 2}_1;{\bf 36})
+\frac{1}{2}({\bf 28},{\bf 1};{\bf 36})+ 10 ({\bf 1},{\bf 1}_2;{\bf 1}) + ({\bf 1},{\bf 1};{\bf 630})$.
In addition, there are the universally present twenty neutral hyper multiplets.

On the orbifold side, as discussed in~\cite{Aldazabal:1997wi} for $T^4/Z_N$ 
cases, 5-branes act as magnetic sources on fixed points thereby shifting
the vacuum energy in the corresponding left-moving non-supersymmetric twist sector
as well as entering the modular invariance constraint while the right-moving
supersymmetric sector is not affected. 5-branes away from the 
fixed points provide the same matter states as in the smooth $K3$ case, but
at fixed points gauge enhancements can occur, as e.g. $Sp(2N_a) \rightarrow U(2N_a)$
for the S-dual D5-branes in the Type I orbifold of~\cite{Gimon:1996rq}.

In~\cite{Aldazabal:1997wi}, the $T^4/Z_3$ embedding with $\vec{V}=\frac{1}{3}(1,1,0_{14})$
and 18 5-branes is listed with perturbative gauge group $SO(28) \times SU(2)_{\times U(1)}$.
Omitting the $U(1)$ charges, the untwisted spectrum is identical to that of the standard embedding 
in table~\ref{tab:Orbifold_SE},
$({\bf 28},{\bf 2}) + 3 ({\bf 1})$, but the twisted spectrum contains only 
$9({\bf 1}) + 18 ({\bf 1}^{\star})$. Adding the states charged under $Sp(36)$, the massless
spectrum of the smooth example with 5-branes is recovered except for the 
massless $U(1)$ at the orbifold point.

\section{Some results in four dimensions}
\label{sec:HeteroticCY} 

The discussion of the stringy consistency conditions and generic spectrum
has been presented for $CY_n$ folds for any $n$. The concrete form of the 
index counting chiral states depends on the dimension and is for $CY_3$-folds 
given by 
\begin{equation}
\chi(CY_3, W ) = \int_{CY_3} \left({\rm ch}_3(W) + \frac{1}{12}c_2(CY_3) c_1(W) \right). 
\end{equation}
Furthermore, as anticipated above, the primitivity condition
on supersymmetric bundles receives a 1-loop correction~\cite{Blumenhagen:2005pm}
\begin{equation}
\int_{CY_3} J^2 \wedge {\rm tr} \bar{F}_i - \frac{2 \, g_s^2}{3}  \int_{CY_3} 
\left( {\rm tr} \bar{F}^3_i - \frac{1}{16} {\rm tr} \bar{F}_i \wedge {\rm tr} \bar{R}^2  \right) =0.
\end{equation}
This is the S-dual generalisation~\cite{Blumenhagen:2005zh}
to non-Abelian bundles on curved backgrounds of the `MMMS' calibration 
condition~\cite{Marino:1999af}
for supersymmetric D9-branes.

The holomorphicity condition is now trivially fulfilled, but the 
requirement that the 1-loop corrected gauge kinetic function is real 
gives a new constraint,
\begin{equation}
N_i \int_{CY_3} J^3- 6 \, g_s^2  \int_{CY_3} J \wedge 
\left( {\rm tr} \bar{F}^2_i - \frac{N_i}{48} \, {\rm  tr} \bar{R}^2  \right) > 0.
\end{equation}

The mass terms from couplings to wrapped modes of the antisymmetric tensor are as given above.
In addition, the coupling of the reduction of the antisymmetric tensor with two external indices,
$B^{(2)} =  b^{(2)}_0 + \ldots$,
also provides a mass coupling,
\begin{equation}
S^0_{mass} \sim \frac{1}{3}\sum_i \int_{M^4} b^{(2)}_0 \wedge {\rm tr}F_i \left( {\rm tr} \bar{F}^3_i - \frac{1}{16}{\rm tr}\bar{F}_i \, {\rm tr} \bar{R}^2  \right).
\end{equation}

The $U(n)$ bundles provide new possibilities for model building with some first results reported 
in~\cite{Blumenhagen:2005pm,Blumenhagen:2005zg} for Complete Intersection and elliptically fibered
Calabi Yau threefolds, respectively.

At the orbifold point, the tree level and 1-loop part of the supersymmetry condition vanish
separately since the volume of the exceptional divisors is zero, but the gauge bundle and curvature
have only support there.

\section{Results on $E_8 \times E_8$ compactifications}
\label{sec:E8_results}

The Bianchi identity, K-theory constraint and supersymmetry condition at tree level
are the same as for the $SO(32)$ case presented above. In four dimensions, however,
the 1-loop contribution to the supersymmetry condition differs. On the one hand,
it involves all bundles inside the same $E_8$ factor, on the
other hand, also the 5-brane positions enter. This is in contrast to the $SO(32)$ case
where the supersymmetry conditions on bundles and 5-branes are decoupled.
In six dimensions, $E_8 \times E_8$ 5-branes along the non-compact directions 
provide tensor multiplets, whereas in four dimensions, space-time filling 5-branes wrap
compact 2-cycles and provide $U(1)$ gauge groups.

The mass terms from wrapped antisymmetric tensor modes have the same formal expression~(\ref{eq:GS_mass}) 
as for the $SO(32)$ case, but in four dimensional compactifications the coupling to
the universal $b^{(2)}_0$ has a different shape.

The differences between the $E_8 \times E_8$  and $SO(32)$ compactifications can be 
traced back to the fact that $E_8$ has no fourth order Casimir. Moreover, while
$U(1)$ and $SU(n)$ groups arise naturally in breakings of $E_8$, $U(n)$ bundles
can only be implemented in a very restricted way, e.g. as $U(n_1) \times U(n_2)$
with $c_1(V_{n_1}) = - c_1(V_{n_2})$.

More details about smooth six dimensional $E_8 \times E_8$ constructions on $K3$ 
with $U(n)$ bundles are given in~\cite{Honecker:2006dt,Honecker:2006qz}, 
the corresponding supersymmetry conditions on $CY_3$ are given in~\cite{Blumenhagen:2005ga}
without and in~\cite{Blumenhagen:2006ux} with 5-branes.

%
% BibTeX users please use
% \bibliographystyle{}
% \bibliography{}

\end{document}